\documentclass[conference]{IEEEtran}

\IEEEoverridecommandlockouts

\usepackage{cite}

\usepackage{amsmath,amssymb,amsfonts,mathrsfs,bm}

\usepackage{graphicx}
\usepackage{subfig}
\graphicspath{{./figures/}}

\usepackage{tabularx}
\usepackage{multirow}
\usepackage{booktabs}

\makeatletter
  \newif\if@restonecol
\makeatother

\usepackage[linesnumbered,ruled,lined]{algorithm2e}
\usepackage{algpseudocode}

\usepackage{flushend}

\usepackage{scalerel}
\usepackage{tikz}
\usetikzlibrary{svg.path}

\definecolor{orcidlogocol}{HTML}{A6CE39}
\tikzset{
  orcidlogo/.pic={
    \fill[orcidlogocol] svg{M256,128c0,70.7-57.3,128-128,128C57.3,256,0,198.7,0,128C0,57.3,57.3,0,128,0C198.7,0,256,57.3,256,128z};
    \fill[white] svg{M86.3,186.2H70.9V79.1h15.4v48.4V186.2z}
                 svg{M108.9,79.1h41.6c39.6,0,57,28.3,57,53.6c0,27.5-21.5,53.6-56.8,53.6h-41.8V79.1z M124.3,172.4h24.5c34.9,0,42.9-26.5,42.9-39.7c0-21.5-13.7-39.7-43.7-39.7h-23.7V172.4z}
                 svg{M88.7,56.8c0,5.5-4.5,10.1-10.1,10.1c-5.6,0-10.1-4.6-10.1-10.1c0-5.6,4.5-10.1,10.1-10.1C84.2,46.7,88.7,51.3,88.7,56.8z};
  }
}

\newcommand\orcidicon[1]{\href{https://orcid.org/#1}{\mbox{\scalerel*{
\begin{tikzpicture}[yscale=-1,transform shape]
\pic{orcidlogo};
\end{tikzpicture}
}{|}}}}

\usepackage[colorlinks,allcolors=black]{hyperref} 

\title{\vspace{0.25in} \LARGE \bf Few-Shot Scenario Testing for Autonomous Vehicles\\Based on Neighborhood Coverage and Similarity}

\author{Shu Li\textsuperscript{\orcidicon{0009-0001-7579-1383}},
Jingxuan Yang\textsuperscript{\orcidicon{0000-0001-9798-7347}},
Honglin He\textsuperscript{\orcidicon{0000-0003-4673-5283}}, 
Yi Zhang\textsuperscript{\orcidicon{0000-0001-5526-866X}}, \IEEEmembership{Senior Member,~IEEE}, \\ 
Jianming Hu\textsuperscript{\orcidicon{0000-0001-8065-7309}}, \IEEEmembership{Senior Member,~IEEE}, and 
Shuo Feng\textsuperscript{\orcidicon{0000-0002-2117-4427}}, \IEEEmembership{Member,~IEEE}%
\thanks{This work is supported by Beijing Nova Program under Grant 20230484259. We acknowledge the Wuhan East Lake High-Tech Development Zone (also known as the Optics Valley of China, or OVC) National Comprehensive Experimental Base for Governance of Intelligent Society and Hebei Expressway Group Limited. \textit{(Corresponding author: Shuo Feng.)}}%
\thanks{Shu Li is with the PKU-WUHAN Institute for Artificial Intelligence, Wuhan 430074, China, and also with the Department of Automation, Tsinghua University, Beijing 100084, China (email: li-s23@mails.tsinghua.edu.cn).}%
\thanks{Jingxuan Yang, Honglin He and Jianming Hu are with the Department of Automation, Tsinghua University, Beijing 100084, China (email: yangjx20@mails.tsinghua.edu.cn, hehl21@mails.tsinghua.edu.cn, hujm@tsinghua.edu.cn).}%
\thanks{Yi Zhang is with the Department of Automation, Beijing National Research Center for Information Science and Technology (BNRist), Tsinghua University, Beijing 100084, China, and also with the Tsinghua-Berkeley Shenzhen Institute (TBSI), Shenzhen 518055, China (e-mail: zhyi@mail.tsinghua.edu.cn).}%
\thanks{Shuo Feng is with the Department of Automation, Beijing National Research Center for Information Science and Technology (BNRist), Tsinghua University, Beijing 100084, China (e-mail: fshuo@tsinghua.edu.cn).}
}

\begin{document}

\maketitle
\thispagestyle{empty}
\pagestyle{empty}

\begin{abstract}
  Testing and evaluating the safety performance of autonomous vehicles (AVs) is essential before the large-scale deployment. Practically, the number of testing scenarios permissible for a specific AV is severely limited by tight constraints on testing budgets and time. With the restrictions imposed by strictly restricted numbers of tests, existing testing methods often lead to significant uncertainty or difficulty to quantifying evaluation results. In this paper, we formulate this problem for the first time the ``few-shot testing" (FST) problem and propose a systematic framework to address this challenge. 
  To alleviate the considerable uncertainty inherent in a small testing scenario set, we frame the FST problem as an optimization problem and search for the testing scenario set based on neighborhood coverage and similarity. Specifically, under the guidance of better generalization ability of the testing scenario set on AVs, we dynamically adjust this set and the contribution of each testing scenario to the evaluation result based on coverage, leveraging the prior information of surrogate models (SMs). With certain hypotheses on SMs, a theoretical upper bound of evaluation error is established to verify the sufficiency of evaluation accuracy within the given limited number of tests. The experiment results on cut-in scenarios demonstrate a notable reduction in evaluation error and variance of our method compared to conventional testing methods, especially for situations with a strict limit on the number of scenarios.
  
\end{abstract}


\section{Introduction}\label{sec:intro}
The fast development and experimental application of high-level autonomous vehicles (AVs) on open road have brought about the crucial need and emerging research interests\cite{li2018artificial, li2019parallel, li2022features} on testing and evaluating their safety performance to facilitate the large-scale deployment. However, with seemingly endless traffic scenarios in real world\cite{riedmaier2020survey} and the low efficiency of testing rare events (e.g. crashes)\cite{liu2022curse}, the available testing budget is far from meeting the practical requirements. Consequently, how to generate reliable testing and evaluation results within the confines of a restricted limit of testing number becomes a significant problem.

In practical terms, the acceptable budget of testing specific AV model can be restricted within an extremely small limit. For third-party testing organizations and governmental bodies, the generation of an extensive array of testing scenarios for all potential AV models, particularly during open road testing, is not pragmatic.
Besides, with the rapid iterative development of autonomous driving technique, conducting a thorough evaluation of AV performance within the research and development cycle becomes increasingly infeasible. Consequently, there exists a compelling need to generate quickly the accurate evaluation results of AV using the smallest possible testing numbers or within a strictly limited set of scenarios. Moreover, the quantitative and explainable results are needed as a foundational benchmark for comparing the performance of diverse AVs, which create additional difficulties for testing with limited costs. Remarkably, we term this problem the ``few-shot testing" (FST) problem in this paper, marking the first instance of defining and addressing this specific issue to the best of our knowledge.

Although a lot of efforts have been made to search for a smaller testing scenario set or accelerated tests, the problem of FST remains unsolved. As a practical method, some autonomous driving companies keep a scenario set from logged data or expert knowledge to verify the reliability of their AVs before on-road deployment\cite{webb2020waymo}. Searching for critical scenarios is a commonly used scheme to generate a smaller testing scenario set\cite{zhang2022finding}. Based on knowledge\cite{bagschik2018ontology}, scenario clustering\cite{kruber2018unsupervised, kruber2019unsupervised}, scenario coverage\cite{weissensteiner2023operational}, optimization strategy\cite{duan2020test, klischat2019generating} or some well-designed models\cite{li2021scegene, li2016intelligence, ge2024task, zhang2018accelerated}, many methods are capable of generating a representative scenario set with certain risks. However, the efficiency of these methods is usually measured by the proportion of critical scenarios or some pre-defined reasonable criteria. The quantitative result for the performance index of AV is hard to be acquired.

Statistical methods represent an effective approach for quantifying the performance index of AV model while generating critical scenarios to accelerate the testing process\cite{feng2020part1, feng2020part2, feng2020testing, zhao2016accelerated, zhao2017accelerated, yang2022adaptive, yang2023adaptive}. Based on naturalistic driving data (NDD) and naturalistic driving environment (NDE), the performance of AVs can be estimated with a critical distribution. However, although unbiased results can be acquired through sampling, many critical scenarios may be similar or reduplicative. Additionally, controlling the testing variance in a smaller test set becomes challenging due to uncertainty, thereby diminishing the effectiveness of these methods in FST scenarios.

In this paper, we propose systematically the FST framework to tackle the problem of quantifying the performance of AV with smallest possible or strictly limited numbers of test. Meanwhile, A testing error bound with regard to the restricted number of testing scenarios can be derived to verify the accuracy of FST method. In order to maintain the theoretical advantages and quantify the performance indices of AV, we generate testing scenarios based on NDD and NDE. Furthermore, to mitigate the large uncertainty of NDE-based statistical methods associated with the restricted numbers of test, we frame the FST problem as an optimization problem and iteratively search for FST scenario set for an accurate and reliable testing and evaluation result. In the context of fixed few-shot scenarios and an unknown AV model before testing, the essence of FST method lies not in the direct selection of critical scenarios but in identifying a set of scenarios with optimal generalization ability based on SMs. This is similar to the generalization ability concept employed in classic few-shot learning methods \cite{finn2017model, koch2015siamese}.

Following the basic idea of FST framework, we introduce a dynamic neighborhood coverage set and a similarity measurement for scenarios in the test set. According to the limit on numbers of testing scenarios and specific scenario samples, the confidence and coverage of scenarios are dynamically adjusted in the optimization process under the guidance of the generalization ability or the estimation error bound. With the gradient descent method, a fixed small set of scenarios can be selected with a minimized upper bound of error. The performance index of real AV model could be quantified with definite accuracy with some hypothesis on SMs.
In order to validate the FST method, we conducted a case study in the cut-in scenario. Compared with commonly used testing methods, the average error and variance of FST is significantly smaller. Notably, we find that the accuracy of FST method is less impaired with a smaller size of testing set compared with the other methods.

The remainder of the paper is organized as follows: Section \ref{sec:formulation} and \ref{sec:method} provide the basic formulation for the FST problem and a comprehensive presentation of our detailed FST framework; Section \ref{sec:case_study} validates the effectiveness of proposed FST method in the cut-in scenario; Finally section \ref{sec:conclusion} concludes the paper.

\section{Problem Formulation}\label{sec:formulation}

\subsection{Performance Index Quantification with NDE}

The modeling of scenarios and environments constitutes a fundamental aspect of autonomous vehicle (AV) testing and evaluation. As mentioned above, a commonly used method for driving environment modeling is the NDE based on NDD, which is defined as follows.

Let $x$ denote a scenario including all spatio-temporal varibles relevant to testing requirements (e.g. position, velocity of vehicles at a specific moment or in a period of time) and we have
\begin{equation}
  x\in\mathcal{X},
\end{equation}
where $\mathcal{X}$ is the set of all possible scenarios. With statistics given by NDD, the probability measure of specific scenario can be derived as $p(x), x\in\mathcal{X}$, which is the exposure frequency of the scenario in real world. Define $X$ as the corresponding random variable and then we have $X \sim p(x)$. Provided that $A$ is the event of interest during testing (usually crashes for AVs), the performance index of AVs can be calculated as
\begin{equation}
\begin{aligned}
    \mu = P(A) &= \mathbb{E}_p[P(A|X)]\\
               &= \sum_{x\in \mathcal{X}}{P(A|x)p(x)},
\end{aligned}
\end{equation}
where $\mu = P(A)$ is the performance index with respect to event $A$ in NDE and $P(A|x)$ is the performance measure of the vehicle under test in the scenario $x$.

\subsection{Statistical Testing Problem}
Classic statistical methods evaluate the performance index of AV by sampling a set of scenarios $\mathcal{X}_{s} \triangleq \{x_i\in\mathcal{X}, i=1,2,...\}$. Then $\mu$ can be estimated by testing AV performances on $x_i$ as $P(A|x_i)$.
Crude Monte Carlo (CMC)\cite{mcbook} method is extensively adopted to test AVs in NDE. The strategy for CMC to quantify $\mu$ is
\begin{equation}\label{eq:cmc}
    \Tilde{\mu}_{\rm{CMC}} = \frac{1}{n}\sum_{i=1}^{n}{P(A|x_i)}, X_i\sim p(x),
\end{equation}
where $\mathcal{X}_{s}$ is generated from natural distribution $p(x)$ of NDE and $n$ could increase dynamically throughout the testing procedure. When $A$ denotes crashes, the crash rate of AV is generally a tiny value and CMC faces the problem of ``curse of rarity", which results in extremely low efficiency and accuracy\cite{liu2022curse}. Consequently, it is practically impossible to evaluate AVs with CMC coupled with a small number of tests.

Towards addressing this issue, importance sampling (IS) is proposed to accelerate the testing process\cite{zhao2016accelerated, zhao2017accelerated, feng2020part1}, the strategy for IS to quantify the crash rate is
\begin{equation}\label{eq:is}
    \Tilde{\mu}_{\rm{IS}} = \frac{1}{n}\sum_{i=1}^{n}{P(A|x_i)\frac{p(x_i)}{q(x_i)}}, X_i\sim q(x),
\end{equation}
where $\mathcal{X}_{s}$ is generated from an important distribution $q(x)$. The testing efficiency of IS can be much higher than CMC under ideal condition. However, in practice the estimation result of IS could not converge to an accurate value for a strictly limited number of tests.

\subsection{Few-Shot Testing Problem}\label{sec:formulation_fst}
In this paper, we focus on the situation where a limited size of scenario set $\mathcal{X}_{s}$ is given as $n$, denoting $\mathcal{X}_{s,n}\triangleq \{x_i\in\mathcal{X}, i=1,2,...,n\}$. As $n$ is a small number (e.g. $n=5, 10$), our primary emphasis is on the estimation error caused by specific testing set $\mathcal{X}_{s,n}$ and the goal is to minimize the evaluation error on specific AV model, that is
\begin{equation}\label{eq:err_1}
    \mathop{\min}_{\mathcal{X}_{s,n}}E = |\Tilde{\mu}-\mu|,
\end{equation}
where $\Tilde{\mu}$ is the estimation result. Similar to other testing methods, we obtain $P(A|x_i)$ by testing AV performance of scenarios $x_i$. Additionally, we extend the estimation strategy Eq.~(\ref{eq:cmc}-\ref{eq:is}) to a general form
\begin{equation}\label{eq:est_general}
    \Tilde{\mu} = f(P(A|x_1), ..., P(A|x_n)),
\end{equation}
where $f$ is any possible estimation strategy leveraging $n$ AV testing performances $P(A|x_i), i=1,...,n$. Combining Eq.~(\ref{eq:err_1}) with Eq.~(\ref{eq:est_general}), we have the standard form for few-shot testing problem as
\begin{equation}\label{eq:formulation}
    \mathop{\min}_{\mathcal{X}_{s,n}}E = |f(P(A|x_1), ..., P(A|x_n))-\mu|.
\end{equation}

Statistical testing methods like CMC and IS have theoretical benefits that the estimation is unbiased. However, as scenarios are sampled from probabilistic distribution, the uncertainty is hard to be controlled with few testing numbers and the estimation errors is possible to be a large value. In Eq.~(\ref{eq:formulation}), all relevant notations are fixed value but not random variables, so the uncertainty of sampling is eliminated.

Towards solving the problem in Eq.~(\ref{eq:formulation}), there are still several challenges remaining:

(1) The estimation function $f$ is extremely flexible, which calls for a concrete and effective estimation strategy; 

(2) As $P(A|x_i)$ is obtained for specific AV model and $\mu$ is the ground truth of performance index for this AV, the optimal solution of Eq.~(\ref{eq:formulation}) is relevant to the unknown AV model under test; 

(3) If we have an approximation of $P(A|x_i)$ or $\mu$ before testing, the unknown gap between real AV model and the approximation may directly affect the estimation error and result in a low accuracy.

\section{Few-Shot Testing Framework}\label{sec:method}

\subsection{Upper Bound of Estimation Error}

AV plays an important role in the formulation (\ref{eq:formulation}). Because we do not know the information of AV model before testing, we use the surrogate models (SMs) to represent for the prior information of AV. Note that the generated FST scenario set $\mathcal{X}_{s,n}$ is fixed after the optimization is finished, but there are diverse possible AV models for test. Because the error between SMs and AVs under test may directly affect the estimation error, we suppose the information of SMs form a set $\mathcal{X}_{s,n}$. For each possible SM $m_i\in\mathcal{M}$, the performance measure $P_i(A|x)$ could be tested and acquired. Subsequently, under the hypothesis that real AV model satisfies $m^*\in\mathcal{M}$, the fixed error can be further written as an error bound
\begin{equation}\label{eq:err_2}
    E \leq \mathop{\max}_{m_i\in\mathcal{M}} |\Tilde{\mu}_i-\mu_i|,
\end{equation}
where $\Tilde{\mu}_i$ and $\mu_i$ is the crash rate and estimation of crash rate of $m_i$, respectively:
\begin{equation}
    \Tilde{\mu}_i=f(P_i(A|x_1), ..., P_i(A|x_n)),\ \mu_i=\mathbb{E}_p[P_i(A|X)].
\end{equation}
The hypothesis $m^*\in\mathcal{M}$ is not such strong because we can expand the set $\mathcal{M}$ by introducing noises so that the real AV model can be covered. With a more deterministic SM set $\mathcal{M}$, the relationship of Eq.~(\ref{eq:err_2}) is more compact and vice versa. With this scheme, we transform the optimization problem into minimizing a upper bound of error. The set of scenarios $\mathcal{X}_{s,n}$ target at the best generalization ability among AVs and thus ensuring the few-shot accuracy.

However, in case that the unknown information can not be completely covered with certain noises, namely $m^*\notin\mathcal{M}$, the error bound could be extended for further analysis.
Suppose that $m^*$ could be decomposed into $m'+\Delta m$ where $m'\in\mathcal{M}$, by applying Eq.~(\ref{eq:err_2}) in $m'$ we have the extended error bound
\begin{equation}\label{eq:err_3}
\begin{aligned}
    E &\leq \mathop{\max}_{m_i\in\mathcal{M}} \{|\Tilde{\mu}_i-\mu_i|\}
    + |\Delta\Tilde{\mu}-\Delta{\mu}| \\
    &\leq \mathop{\max}_{m_i\in\mathcal{M}} \{|\Tilde{\mu}_i-\mu_i|\}
    + |\Delta\Tilde{\mu}| + |\Delta{\mu}|.
\end{aligned}
\end{equation}
With these upper bounds of estimation error, the estimation strategy for $\Tilde{\mu}$ could be designed.

\subsection{Coverage and Similarity-Based Estimation Strategy}
As stated in section \ref{sec:formulation_fst}, the form of estimation function $f$ is extremely flexible. 
In this paper, we deal with $f$ using a weighted sum of testing results on all scenarios as
\begin{equation}\label{eq:est_fst}
    f(P(A|x_1),...,P(A|x_n)) = \sum_{i=1}^{n}{P(A|x_i)w(x_i;\mathcal{X}_{s,n})},
\end{equation}
where $w(x_i;\mathcal{X}_{s,n})$ is the weight function of each scenario sample $x_i$ and we suppose that $\sum_{i=1}^{n}{w(x_i;\mathcal{X}_{s,n})}=1$. 
Note from Eq.~(\ref{eq:est_fst}) that there is $w(x_i;\mathcal{X}_{s,n})=1/n$ for CMC method and $w(x_i;\mathcal{X}_{s,n})=p(x_i)/[nq(x_i)]$ for IS method. Traditional weight functions are determined only by certain sample $x_i$ and the size of test set $n$ hence lacking the global perspective. Besides, the $\mathcal{X}_{s,n}$ is sampled based on distributions, which brings about large uncertainty. By making use of a relatively flexible form of weight functions and test set, the few-shot accuracy may be guaranteed.

Now we come to the discussion on the weight function. Following the idea of fully utilizing each valid testing scenario and avoiding redundant information, we want to take advantage of the concept of scenario coverage. The coverage of a scenario is hopeful to be a reliable measurement of its contribution to the evaluation result. Coverage is a commonly used and effective approach to verify the performance of AVs\cite{amersbach2019defining}. Experientially and intuitively, the similarity among scenarios allows us to utilize the performance of AV in a representative scenario as an approximation of its neighborhood. 

Usually the coverage of scenario is deterministic with some pre-determined rules or models\cite{tahir2020coverage, zhao2022automated}. As FST method aims at evaluating the performance of AV under fixed size of test set, the accuracy and expected contribution of each scenario sample for FST should change along with the size of test set. Therefore, we propose a dynamic neighborhood coverage set to adjust the relative coverage of scenarios selected for test dynamically with respect to $\mathcal{X}_{s,n}$. The coverage of one sample $x_i\in\mathcal{X}_{s,n}$ is decided by the similarity between all scenarios in the scenario space and the test set $\mathcal{X}_{s,n}$. We define the coverage set of $x_i$ by
\begin{equation}
    \mathcal{C}(x_i; \mathcal{X}_{s,n})\triangleq\left\{x'|\mathop{\arg\max}_{j=1,...,n}S(x', x_j)=i\right\},
\end{equation}
where $S(x', x)$ is a similarity measurement which can be defined as the inverse of normalized Euclidean norm for simplicity
\begin{equation}
    S(x', x)\triangleq \frac{1}{||x-x'||_2}. 
\end{equation}
By adopting this brief form of similarity measurement, the boundary of coverage sets is actually a separate hyperplane in the scenario space. Then the weight of $x_i$ is computed with the sum of all scenarios in its coverage set weight by exposure frequency
\begin{equation}\label{eq:w}
    w(x_i; \mathcal{X}_{s,n})=\sum_{x\in\mathcal{C}(x_i; \mathcal{X}_{s,n})}p(x).
\end{equation}

With the definition of coverage set and weight function, the estimation $\Tilde{\mu}$ is continuous providing the continuity of scenario space $X, p(x)$ and performance measure function $P(A|x)$. With this property it is easy to prove that for specific AV model $m^*$ and $n\geq 2$, the optimal estimation error $\Tilde{\mu}$ is $0$ by selecting optimal $\mathcal{X}_{s,n}$. This ensures the theoretical optimality of FST method with exactly accurate prior knowledge. Specifically, the real-world scenario variables are usually continuous but the discretization precision may affect the optimality to some extent.

\subsection{Optimizing for a Few-Shot Scenario Set}

Eventually with the information of SMs set $\mathcal{M}$, the gradient descent method is conducted to search for a optimal test set $\mathcal{X}_{s,n}$ and dynamically adjust coverage of samples $x_i$ in each iteration. By directly applying the error upper bound in Eq.~(\ref{eq:err_2}), we write the objective function for optimization as 
\begin{equation}\label{eq:obj_1}
\begin{aligned}
    \mathop{\min}_{\mathcal{X}_{s,n}}\ J(\mathcal{X}_{s,n})& \\
    \mathrm{s.t.}\ J(\mathcal{X}_{s,n})&=\mathop{\max}_{m_i\in\mathcal{M}}|\Tilde{\mu}_i-\mu_i|,
\end{aligned}
\end{equation}
where $\Tilde{\mu}_i$ is calculated with Eq.~(\ref{eq:est_fst})-(\ref{eq:w}).

With the extended upper bound of error in Eq.~(\ref{eq:err_3}), 
we can see that $\Delta{\mu}=\mathbb{E}_p[\Delta P_i(A|X)]$ is the unknown overall gap between real AV model and SM set. This item is irrelevant to the estimation strategy $f$ or scenario samples $\mathcal{X}_{s,n}$ and is impossible to be eliminated. Therefore, we focus on the additional part $|\Delta\Tilde{\mu}|$ in Eq.~(\ref{eq:err_3}), which can be written as
\begin{equation}\label{eq:err_add}
    |\Delta\Tilde{\mu}| = \left|\sum_{i=1}^{n}{\Delta P(A|x_i)w(x_i;\mathcal{X}_{s,n})}\right|.
\end{equation}
$|\Delta\Tilde{\mu}|$ represents for the gap between SM and AV at specific sample $x_i$ and it means that the information of SM at sampled scenarios should be accurate, particularly at significant scenarios with large weight $w(x_i;\mathcal{X}_{s,n})$.

As $P(A|x_i)$ is undiscovered before testing, we propose the coverage fluctuation estimator to approximate the potential error $\Delta P(A|x_i)$ of scenarios in the test set $\mathcal{X}_{s,n}$, defined by
\begin{equation}
\begin{aligned}
    F(x_i; \mathcal{X}_{s,n})\triangleq\frac{\sum_{x\in\mathcal{C}(x_i)}{[P'(A|x)-P'(A|x_i)]p(x)S(x, x_i)}}{\sum_{x\in\mathcal{C}(x_i)}{p(x)S(x, x_i)}}.
\end{aligned}
\end{equation}
This fluctuation estimator is the summation of differences between $x_i$ and scenarios in its coverage set weighted by the similarity measurement and exposure frequency. It is practically effective that if a scenario has significant differences with another one that is identified as similar, the underlying uncertainty should be taken into consideration.

Replacing $\Delta P(A|x_i)$ in Eq.~(\ref{eq:err_add}) with the fluctuation estimator, substituting it in to Eq.~(\ref{eq:obj_2}) and ignoring the constant part, the optimization principle is rewritten as
\begin{equation}\label{eq:obj_2}
\begin{aligned}
    \mathop{\min}_{\mathcal{X}_{s,n}}\ J(\mathcal{X}_{s,n})& \\
    \mathrm{s.t.}\ J(\mathcal{X}_{s,n})&=w_\mathcal{M}\mathop{\max}_{m_i\in\mathcal{M}}\{|\Tilde{\mu}_i-\mu_i|\}\\
    &+\sum_{i=1}^{n}{F(x_i; \mathcal{X}_{s,n})w(x_i;\mathcal{X}_{s,n})}.
\end{aligned}
\end{equation}
Note that $w_\mathcal{M}$ is our confidence parameter on the prior knowledge provided by SMs. If the AV model is predictable with the assistance of SMs, $w_\mathcal{M}$ is supposed to be set to $+\infty$ and the objective function in Eq.~(\ref{eq:obj_2}) degrades into Eq.~(\ref{eq:obj_1}). The gradient descent method is also applicable with this additional form.

By leveraging scenario coverage and similarity, FST method is capable of selecting optimal test set with an upper bound of error. When the unknown gap between AV and SMs can not be neglected, additional error caused by scenarios selected can also be controlled. Because the formulation targets at the optimized testing error given specific hyper-parameter $n$ and utilizes global information to select test set with strongest generalization ability among AV models, FST is possible to tackle the inaccuracy issue with small testing numbers. As the randomness is restricted in a finite range, it can be used to measure whether the testing error is acceptable within fixed number $n$ of scenarios.

\section{Case Study}\label{sec:case_study}

\subsection{Cut-in Scenario}
\begin{figure}[!t]
  \centering
  \includegraphics[width=8.85cm]{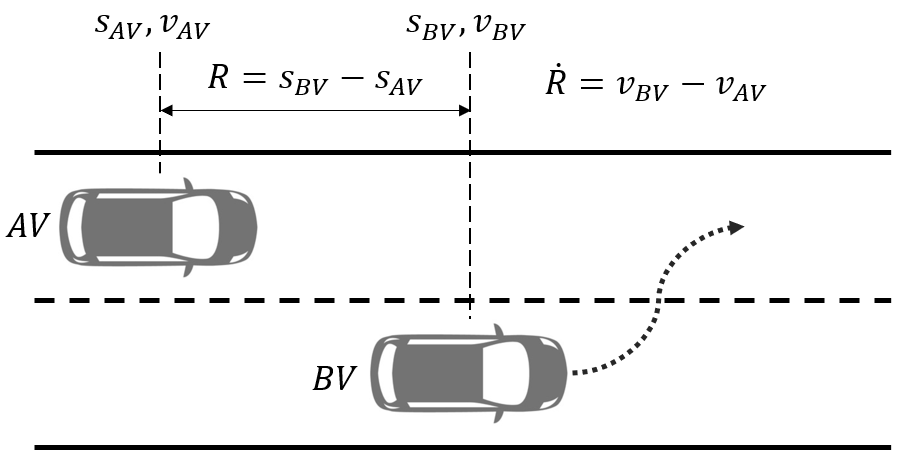}
  \caption{Cut-in scenario.}
  \label{fig:cutin}
\end{figure}

Cut-in is a scenario with high frequency and definite risk in traffic\cite{feng2020part2}, which is adopted in this paper for the testing experiment. As shown in Fig.~\ref{fig:cutin}, the state of cut-in scenario could be simplified as a 2-dimensional variable
\begin{equation}
    x=[R, \dot{R}],
\end{equation}
where $R$ and $\dot{R}$ denote the range and range rate respectively between the background vehicle (BV) and AV at the moment of lane change. Considering the event $A$ as crashes, if the AV fails to speed down to avoid BV in scenario $x$ such that the distance between two vehicles break the threshold $d=s_{BV}-s_{AV}<d_{th}$, the AV would crash with BV. If crashes happen we have $P(A|x)=1$ and otherwise $P(A|x)=0$. By extracting real world driving behaviors from NDD, we construct the exposure frequency $p(x)$ of each scenario variable at cut-in moments. Therefore $P(A)$ denotes the overall crash rate of AV. On the basis of NDD, the scenario space is restricted within
\begin{equation}
    \mathcal{X}=\{[R,\dot{R}]| R\in(0,90], \dot{R}\in[-20, 10]\}.
\end{equation}
By initializing specific cut-in scenario and simulating actions for AV and BV for a sufficient period of time, $P(A|x)$ and crash rate for vehicle models can be tested.

\subsection{Experiment Settings}
\begin{figure}[!t]
  \centering
  \includegraphics[width=8.85cm]{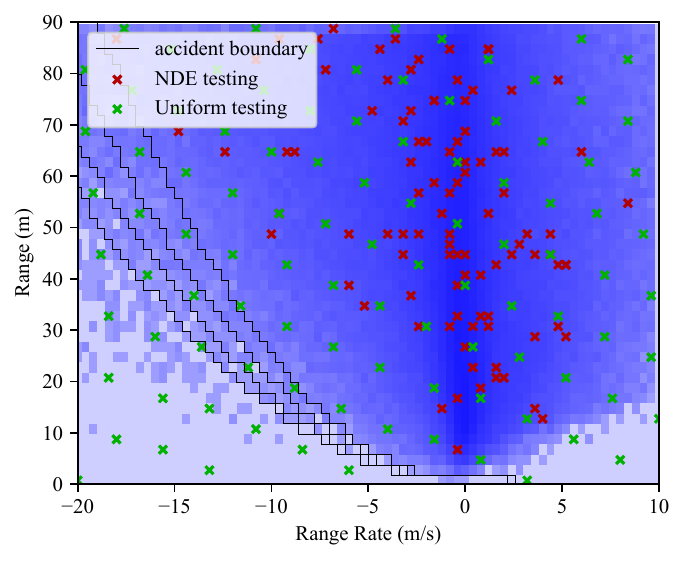}
  \caption{Illustration of basic testing methods and SMs. Scenarios with smaller range and range rate in the left side of boundary would encounter crashes in simulation. The saturability of background demonstrate the exposure frequency in NDD.}
  \label{fig:baseline_samples}
\end{figure}

Intelligent Driving Model (IDM) is a simple and widely-used driving model to imitate human driving behaviors. In order to construct a SM set with diverse human driving manners and tendencies, we use 4 IDMs (denoting $m_1,...,m_4$) with different parameters from conservative to aggressive as the basis and the SM set is a linear combination of these models:
\begin{equation}
    \mathcal{M} = \left\{m'|m'=\sum_{i=1}^{4}{c_im_i},\ c_i\geq 0,\ \forall i,\  \sum_{i=1}^{4}{c_i}=1\right\}.
\end{equation}
The performances of SMs are shown in Fig.~\ref{fig:baseline_samples}. The separate lines are the crash boundaries of 4 SMs between $P(A|x)=0$ and $P(A|x)=1$. Accidents appear in the left side of boundaries with smaller range and range rate and the overall crash rates of SMs vary from $4.6\times 10^{-4}$ to $4.9\times 10^{-3}$. The exposure frequency of scenarios in NDD is illustrated with the saturability of background color. We can see that most scenarios with large exposure frequency $p(x)$ are crash-free. This results in the rarity of crash events.

The AV model for test is set to another group of IDM parameters with the crash rate of $3.0\times 10^{-4}$ and BV keeps a constant velocity after changing lane. As the simulation in our experiment is a complex temporal process, it is hard to fitting the exact crash performance of AV in all scenarios with 4 SMs and we have $m^*\notin\mathcal{M}$. Therefore, we set the confidence parameter $w_\mathcal{M}=1$.

With the property of linear combinations for $\mathcal{M}$, we have
\begin{equation}
    \mathop{\max}_{m'\in\mathcal{M}} |\Tilde{\mu}'-\mu'|=\mathop{\max}_{m_i, i=1,...,4} |\Tilde{\mu}_i-\mu_i|,
\end{equation}
and the optimization objective function can be computed. In order to introduce randomness in FST method, we randomly initialize the test set $\mathcal{X}_{s,n}$ and perform gradient descent to search for optimal set $\mathcal{X}_{s,n}$.

As comparisons, we applied CMC testing in NDE and the uniform sampling method random quasi-Monte Carlo (RQMC)\cite{practicalqmc} to test the AV model. The results of 100 testing scenarios generated by NDE and uniform sampling are also shown in Fig.~\ref{fig:baseline_samples}. Because of the rarity of crash events, almost all scenarios generated in NDE concentrate in unchallenging areas, which make the testing and evaluation result effectless. RQMC selects scenarios uniformly in the scenarios space with randomness. The scenarios information is extracted evenly to evaluate the performace of AV. With a small number of test, scenarios with high risks might be tested.

With the purpose of verifying FST method with diverse sizes of test set and examining the error bound, we set $n=5, 10, 20$ as the restricted numbers of test. With each number of testing scenarios, the baseline methods and FST method were repeated for 100 times.

\subsection{Evaluation Results}

\begin{figure}[!t]
  \centering
  \includegraphics[width=8.85cm]{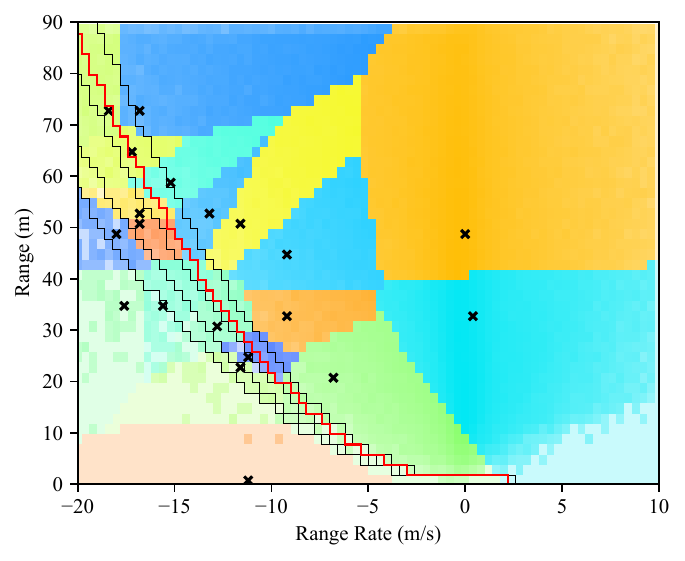}
  \caption{Example of 20 samples and the coverage given by FST method.}
  \label{fig:FST_samples}
\end{figure}

We use an example of 20 scenarios sampled by FST in Fig.~\ref{fig:FST_samples} to illustrate the scenario samples selection and coverage division strategy. The different colors is used to distinguish the coverage set of samples with the saturability hinting the exposure frequency. The crash boundary of AV model under test is shown with the red line. We can see that the coverage set is large for samples in regions where the scenarios are not challenging enough. It means that only a smaller number of tests are sufficient to verify the performance of AVs. In regions where the SMs exhibit disparate performances, more scenarios are sampled and the coverage of each sample is relatively small. This ensures the generalization ability of FST method among distinct AVs in the prior knowledge set $\mathcal{M}$. Moreover, the crash boundary of SMs and AV matches roughly with the boundary of different coverage set, which allows a smaller error bound and higher accuracy for FST method.

\begin{figure}[!t]
  \centering
  \subfloat[$n = 5$]{\includegraphics[width=8.5cm]{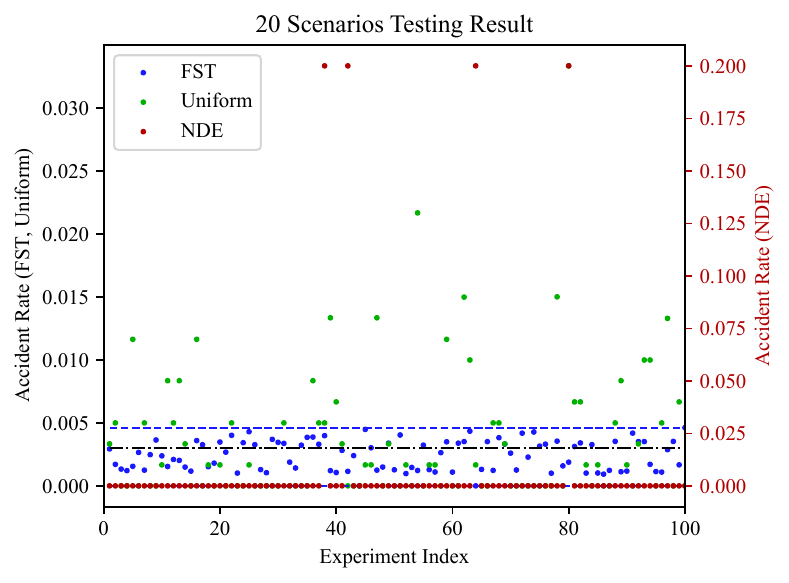}}
  \vfil
  \subfloat[$n = 10$]{\includegraphics[width=8.5cm]{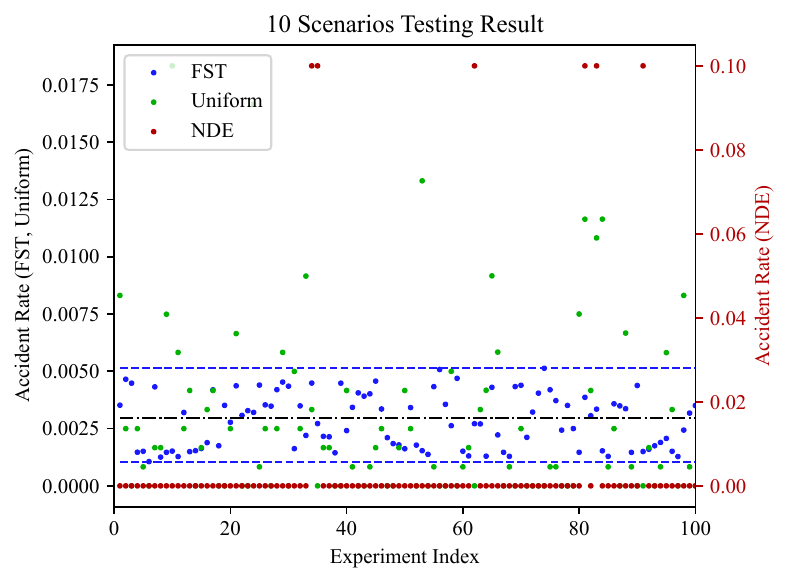}}
  \vfil
  \subfloat[$n = 20$]{\includegraphics[width=8.5cm]{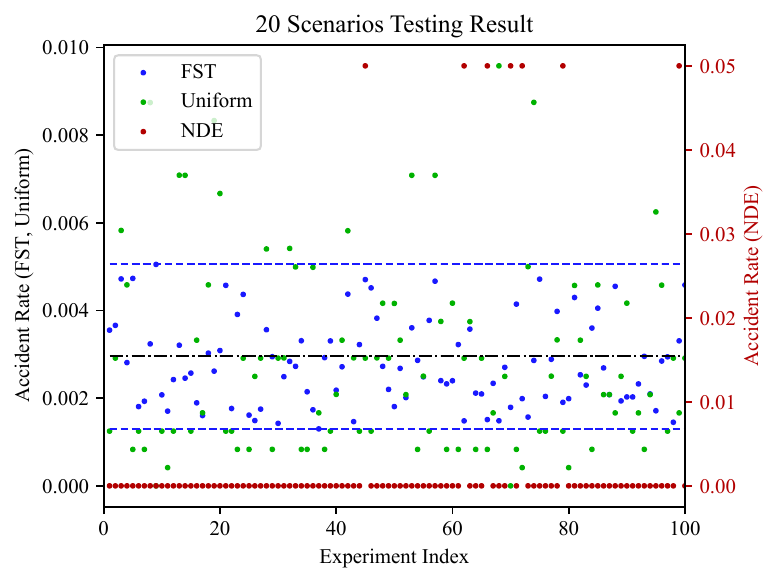}}
  \caption{Testing result of $n = 5, 10, 20$ samples with different methods repeated in 100 experiments.}
  \label{fig:test_res}
\end{figure}

The testing and evaluation results of NDE, uniform sampling and FST are shown in Fig.~\ref{fig:test_res}. For NDE testing, most experiments yield an overall crash rate of $0$ while several experiments produce huge testing errors. It indicates that the accurate performance of AV can hardly be obtained with a small test set in NDE. Uniform sampling method achieves a more accurate evaluation result than NDE but in some experiments the error is large. Compared to the other two methods, the maximum errors of FST method using 5, 10, 20 samples are all bounded in a certain scope (the blue dashed line in Fig.~\ref{fig:test_res} around the real crash rate of AV.

\begin{table}
  \caption{Statistics of testing with $n = 5, 10, 20$ samples}
  \begin{tabular}{ccccccc}
    \toprule
    \multirow{2}{*}{Method} & \multicolumn{3}{c}{Average error $(\times10^{-3})\downarrow$} & \multicolumn{3}{c}{Variance $(\times10^{-6})\downarrow$} \\
    \cmidrule(lr){2-4}\cmidrule(lr){5-7} & $n=5$ & $n=10$ & $n=20$ & $n=5$ & $n=10$ & $n=20$\\
    \midrule
    NDE & $10.7$ & $8.61$ & $6.05$ & $1561$ & $573$ & $163$ \\
    Uniform & $3.74$ & $2.83$ & $1.71$ & $29.2$ & $14.0$ & $4.66$ \\
    FST & $\mathbf{1.11}$ & $\mathbf{1.06}$ & $\mathbf{0.85}$ & $\mathbf{1.68}$ & $\mathbf{1.40}$ & $\mathbf{1.00}$ \\
    \bottomrule
  \end{tabular}
  \label{tab:result}
\end{table}

The average error and variance of 3 methods are shown in TABLE~\ref{tab:result}. The FST method surpasses the common baseline methods significantly in all experiments and indices. Furthermore, the accuracy of FST is less impaired than the other two methods when the number of testing scenarios is smaller. This remarkable feature shows the efficiency and reliability of FST method to testing with restricted budgets or searching for minimum test set.

\section{Conclusion}\label{sec:conclusion}

In this paper, we propose the few-shot testing method to tackle the problem of quantifying AV performance with a strictly limited number of tests. By utilizing the scenario neighborhood coverage and similarity on prior information of models, we iteratively search for a small scenario set to extract scenarios information with the strongest generalization ability. A theoretical error bound can also be estimated with FST to measure whether the accuracy of testing and evaluation result is acceptable for a fixed size of test set. Results show that the proposed method achieves better accuracy than commonly used baseline methods particularly in case that the limit on testing numbers is stricter. In future, more efficient and general designment of scenario weight, coverage and the fluctuation estimator could be developed. Besides, the application of FST method on more complex scenarios is also a potential direction of further studies.

\bibliographystyle{IEEEtran}
\bibliography{IEEEabrv,reference}

\end{document}